\documentclass{emulateapj}

\slugcomment{Manuscript Version Date: \today}

\shorttitle{6 to 37~$\mu$m Imaging of Orion BN/KL with SOFIA}
\shortauthors{De Buizer et al.}

\begin{document}

\title{First Science Observations with SOFIA/FORCAST: 6 to 37~$\mu$m Imaging of Orion BN/KL}

\author{James M. De Buizer\altaffilmark{1}, Mark R. Morris\altaffilmark{2}, E. E. Becklin\altaffilmark{1,2}, Hans Zinnecker\altaffilmark{1}, Terry L. Herter\altaffilmark{3}, Joseph D. Adams\altaffilmark{3}, Ralph Y. Shuping\altaffilmark{4,1}
and William D. Vacca\altaffilmark{1}}

\altaffiltext{1}{SOFIA-USRA, NASA Ames Research Center, MS N211-3, Moffett Field, CA 94035, USA; jdebuizer@sofia.usra.edu}
\altaffiltext{2}{Department of Physics and Astronomy, University of California, Los Angeles, CA 90095-1547, USA}
\altaffiltext{3}{Center for Radiophysics and Space Research, Cornell University, 208 Space Sciences Building, Ithaca, NY 14853 USA}
\altaffiltext{4}{Space Science Institute, 4750 Walnut Street, Boulder, Colorado 80301, USA}

\begin{abstract}
The BN/KL region of the Orion Nebula is the nearest region of high mass star formation in our galaxy. As such, it has been the subject of intense investigation at a variety of wavelengths, which have revealed it to be brightest in the infrared to sub-mm wavelength regime. Using the newly commissioned SOFIA airborne telescope and its 5--40~$\mu$m camera FORCAST, images of the entire BN/KL complex have been acquired. The 31.5 and 37.1~$\mu$m images represent the highest resolution observations ($\lesssim$4$\arcsec$) ever obtained of this region at these wavelengths. These observations reveal that the BN object is not the dominant brightness source in the complex at wavelengths $\geq$31.5~$\mu$m, and that this distinction goes instead to the source IRc4. It was determined from these images and derived dust color temperature maps that IRc4 is also likely to be self-luminous. A new source of emission has also been identified at wavelengths $\geq$31.5~$\mu$m that coincides with the northeastern outflow lobe from the protostellar disk associated with radio source I.

\end{abstract}

\keywords{infrared: ISM --- ISM: individual objects (Orion-BN, Orion Kleinmann-Low) --- stars: massive}

\section{Introduction}

Ever since the discovery of the Becklin-Neugebauer object and the Kleinmann-Low nebula (Becklin \& Neugebauer 1967, Kleinmann \& Low 1967) the region of star birth about 1$\arcmin$ northwest of the Trapezium stars has remained a prominent focus of infrared studies. In part, the attention on BN/KL is due to the luminosity ($\sim$10$^5$ L$_{\sun}$) and proximity (416$\pm$6 pc; Kim et al. 2008) of the region, which suggest it is the nearest location of ongoing high-mass star formation to the Sun. Another reason for the interest in this small complex of very young stars is the mystery of what exactly is happening in this dense, heavily obscured, embedded cluster of infrared sources (Rieke et al. 1973, Lonsdale et al. 1982). Processes including accretion and outflow are occurring, and even perhaps protostellar collisions. Summaries of the variety of observations carried out over the years have been given by Genzel \& Stutzki (1989) and Bally \& Zinnecker (2005). Despite many years of investigation, many questions are only slowly being answered regarding the BN/KL region, including its dynamical history, and which of the so-far identified objects are the dominant sources of luminosity and activity.

In this Letter, we describe new mid-infrared observations of this region taken by the Stratospheric Observatory for Infrared Astronomy (SOFIA). These observations were motivated by the desire to identify the main luminosity sources of the complex at wavelengths greater than 30~$\mu$m for the first time with $\lesssim$4$\arcsec$ resolution. This study represents a follow-up of the lower-resolution far-infrared KAO mapping of the 50 and 100~$\mu$m surface brightness distributions, which peak sharply on the BN/KL region (Werner et al. 1976) with an integrated total FIR luminosity close to 10$^5$ L$_{\sun}$. It also complements recent ground-based 10 to 20~$\mu$m studies of the region (Gezari et al. 1998, Greenhill et al. 2004, Shuping et al. 2004, Okumura et al. 2011), by overlapping and extending the mid-infrared wavelength coverage of the region.

\section{Observations and Data Reduction}

The SOFIA observations of Orion were performed on the nights of 30 November and 3 December 2010 using the FORCAST instrument (Adams et al. 2010). FORCAST is a dual-array mid-infrared camera capable of taking simultaneous images at two wavelengths. The short wavelength camera (SWC) is a 256$\times$256 pixel Si:As array optimized for 5-25 $\mu$m observations; the long wavelength camera (LWC) is a 256$\times$256 pixel Si:Sb array optimized for 25-40 $\mu$m observations. After correction for focal plane distortion, FORCAST effectively samples at 0.768$\arcsec$ pixel$^{-1}$, which yields a 3.4$\arcmin$$\times$3.2$\arcmin$ instantaneous field-of-view.

Observations were obtained in the 6.6, 7.7, 19.7, 31.5, and 37.1~$\mu$m filters at an aircraft altitude of 43000 ft. The chopping secondary of SOFIA was driven at 2 Hz, with a chop throw of 7$\arcmin$ and telescope was nodded 10$\arcmin$ away from the Orion nebula every 30s. The final effective on-source exposure times for the observations were 150s at 6.6 and 7.7~$\mu$m, 450s at 19.7~$\mu$m, 300s at 31.5~$\mu$m, and 420s at 37.1~$\mu$m.

Because all FORCAST data of BN/KL were taken in the dichroic mode, one can determine precise relative astrometry of the two wavelength images that were obtained simultaneously. Data were taken with the following filter pairs: 19.7$\mu$m/37.1$\mu$m, 19.7$\mu$m/31.1$\mu$m, 6.6$\mu$m/37.1$\mu$m, and 7.7$\mu$m/37.1$\mu$m. Therefore the relative astrometry of all filters was determined precisely by bootstrapping from the 19.7$\mu$m/37.1$\mu$m image pair. The relative astrometry between filters is known to better than 0.5 pixels ($\sim$0.38$\arcsec$). All images then had their astrometry absolutely calibrated by assuming that the peak of the BN source at 19.7~$\mu$m is the same as the peak position quoted by Rodriguez et al. (2005) at 7mm (R.A.[J2000]=05$^h$35$^m$14$\fs$119, Dec[J2000]=-05$\arcdeg$22$\arcmin$22$\farcs$785).

The 19.7, 31.5 and 37.1~$\mu$m data were flux calibrated using $\mu$ Cep. Adopted flux density values of 970, 397 and 280 Jy, respectively, were used. The 6.6~$\mu$m data were flux calibrated using $\mu$ UMa, which was taken to be 204 Jy, and the 7.7~$\mu$m data were flux calibrated using $\beta$ UMi, which was taken to be 238 Jy. All of these flux values were derived by multiplying the spectrum of each star by the filter, telescope, and atmospheric throughputs and integrating over the passbands. The calculated integrated flux densities for several sources in the BN/KL region are tabulated in Table 1. It is estimated that the errors in the flux density values derived in this work are 3$\sigma$=20\% (see Herter et al. 2012 in this volume).

The images at 31.5 and 37.1~$\mu$m were deconvolved using the maximum likelihood method (Richardson 1972; Lucy 1974). Like all deconvolution methods, knowledge of the point-spread function (PSF) of an unresolved source is needed at each wavelength. Since the final images were co-additions of data across different observing conditions and flights, the true PSFs of the final images are unknown. Using standard stars observed throughout several flights, an average FWHM for each wavelength was determined. Then artificially generated PSFs (an Airy pattern calculated from the wavelength, telescope diameter, and central obscuration diameter) were constructed and convolved with a gaussian to achieve PSFs with FWHMs that equalled the measured average FWHMs of the standard stars. These idealized PSFs were then used in the deconvolution procedure. 
The deconvolution routine was stopped at 60 iterations for the 31.5 and  37.1~$\mu$m images. The deconvolved images compare favorably to simple unsharp masking of the original images, and thus the substructures revealed in the deconvolved images are believed to be reliable.

\section{Analysis and Results}

Images from the data taken of Orion BN/KL at four of the five wavelengths observed are shown in Figure 1. The 6.6~$\mu$m data are not shown because they had low signal-to-noise and suffered from poor image quality. The locations of several of the well-known ``IRc'' sources in the region are marked in Figure 1. The first instance of the IRc (``IR clump'') nomenclature is by Rieke et al. (1973), who found five IR peaks of emission at 5--21 $\mu$m, with IRc1 being BN itself. Gezari et al. (1998) extended this nomenclature to IRc18 in the mid-IR, and all of these IRc sources are labeled in Figure 1. We also label for context the locations of the well-known near IR source ``n'' and radio source ``I'', which are the subjects of many recent studies.

From these images it can be seen that BN and IRc2 dominate the emission in the region at 7.7~$\mu$m. At longer wavelengths, IRc4, IRc3, and IRc7 appear to dominate. IRc8, IRc13, IRc14 and IRc18 appear as distinct sources only at 19.7~$\mu$m.  Also, the region as a whole has a similar morphology at 31.5 and 37.1~$\mu$m, likely due to the combination of extended environmental dust emission coming to prominence and the decreased resolution at these longer wavelengths.

The source peak positions labeled in Figure 1 are given in Table 1 relative to the measured peak of BN. For IRc3, IRc4, IRc6, IRc7, IRc8, IRc13, IRc14, IRc18, and SOF1 the offsets given were measured from the 19.7~$\mu$m data shown in Figure 1. For the rest of the sources, where there are no observed definitive peaks, the offsets come from Gezari et al. (1998). The position of source I comes from Rodriguez et al. (2005).

\subsection{Dust Color Temperature and Emission Optical Depth Maps}

Using the astrometrically registered images, dust color temperature (T$_c$) and emission optical depth ($\tau$$_e$) maps were constructed for the region using the 19.7/31.5~$\mu$m and 31.5/37.1~$\mu$m image pairs. This was accomplished by first convolving the shorter wavelength image with a gaussian of sufficient width to yield the same resolution as the longer wavelength image, as determined from standard star observations.

For a given pair of images, the surface brightnesses in each spatially matched pixel were iteratively solved for $\tau$$_e$ (evaluated at the shorter wavelength) and T$_c$ under the assumption of blackbody emission from an absorbing and emitting cloud. These calculations used the relationship $\tau$$_{19.7\micron}$/$\tau$$_{31.5\micron}$ = 3.7 and $\tau$$_{31.5\micron}$/$\tau$$_{37.1\micron}$ = 1.7 from the extinction law of Mathis (1990). This calculation of T$_c$ and $\tau$$_e$ only refers to the layers responsible for the emission at these wavelengths, if the region is optically thick. The procedure yielded values of $\tau$$_e$ greater than 1 across the whole region, signifying that the dust here is indeed near the optically thick limit throughout.

The 19.7/31.5~$\mu$m and 31.5/37.1~$\mu$m color temperature maps for the Orion BN/KL region are shown in Figure 2. From the 19.7/31.5~$\mu$m color temperature map it can be seen that there are two main dust temperature peaks coincident with mid-infrared flux peaks at BN and IRc4. This is a strong indication that these sources are internally heated by a stellar source or sources. However, because of the relative decrease in flux of BN at longer wavelengths, the only definitive temperature peak in the 31.5/37.1~$\mu$m color temperature map is at the location of IRc4.

\subsection{Spectral Energy Distributions and Luminosities}

Since the two most likely internally heated sources in the region are BN and IRc4, the flux densities derived from the observations presented here were combined with flux density measurements from Okumura et al. (2011)\footnote{Okumura et al. (2011) published only dereddened flux densities. To compare with the raw photometry of the FORCAST data, the flux densities in this table were kindly provided in a private communication by S. Okumura. For BN these values are 310, 280, 170, 200, 380, 420, 510, 900 and 1100 Jy at 7.8, 8.8, 9.7, 10.5, 11.7, 12.4, 18.5, 20.8 and 24.8~$\mu$m. For IRc4 these values are 13, 5.5, 3.8, 11, 81, 170, 620, 1500 and 2700 Jy at those same wavelengths.} and Dougados et al. (1993) to create SEDs, shown in Figure 3. The use of photometry data from Spitzer is notably absent, as the images are saturated in the BN/KL region. SEDs were fit with the SED model fitter of Robitaille et al. (2007), which assumes a single central stellar source with different combinations of axisymmetric circumstellar disks, infalling flattened envelopes and outflow cavities. The fitting procedure interpolates the model fluxes to the apertures used in the measurements (given in Table 1 for the FORCAST data), scaling them to a given distance range of a source, which in this case is taken to be 416$\pm$6 pc (Kim et al. 2008). It also takes into account the interstellar extinction to the source, A$_V$, which was taken from the work of Gezari et al. (1998) to be 10$\pm$5 for BN and 40$\pm$3 for IRc4. These models yield estimates of physical parameters, which are highly degenerate in general. This, combined with the inhomogeneous nature of the data used in the SEDs (i.e., not all simultaneously obtained, different wavelength and facility dependent resolutions, etc.), underscore the uncertainties of the derived physical parameters. Even so, the models are still relatively good at estimating the total luminosity of sources, which is the main motivation for using them here.

Where is the bulk of the luminosity coming from in the BN/KL region? In Figure 3, the top ten best model fits to the SED data are shown for the two brightest mid-infrared objects: BN and IRc4. According to these models the bolometric luminosity for BN is in the range of 0.8--2.1$\times$10$^4$ L$_{\sun}$, with a mean value of L(BN)$_{bol}$=1.3$\times$10$^4$ L$_{\sun}$. For IRc4 the models give a range of 1.5-2.9$\times$10$^4$ L$_{\sun}$ with a mean value of L(IRc4)$_{bol}$=2.1$\times$10$^4$ L$_{\sun}$. This means that BN and IRc4 alone account for 25--50\% of the total 10$^5$ L$_{\sun}$ for the entire BN/KL region.

\section{Discussion of Individual Sources}

\subsection{Source BN}

The northernmost source in the BN/KL complex, BN itself, appears to be in some ways unrelated to the phenomena of the rest of the KL region. The polarization maps from Simpson et al. (2006) show that BN is responsible for the heating of its immediate environment; however it does not appear to be influencing the heating or illumination of the rest of the sources seen in the infrared in the KL region.

Robberto et al. (2005) point out that from 2 to 20~$\mu$m BN is the brightest peak in the area. However, the data here show BN is not the dominant source at longer thermal infrared wavelengths. Wynn-Williams et al. (1984) first showed that it is less bright than IRc4, IRc7, and IRc3 at 30~$\mu$m. It is barely resolved as a distinct source in the deconvolved 31.5~$\mu$m image, and in the deconvolved 37.1~$\mu$m image it can no longer be distinguished as a separate source (Figure 1). In fact, BN has a remarkably flat SED (Figure 3), with an almost constant value of $\lambda$F$_{\lambda}$ from 3 to 37~$\mu$m.

\subsection{Sources IRc3, IRc4 and IRc7}

These three sources constitute the bulk of the emission seen at 31.5 and 37.1~$\mu$m. Like most IR sources in the KL region, there has been some speculation on the nature of these objects and whether they are simply knots of dust in the complex nebulosity or embedded stars. For example, Zuckerman et al. (1981) argued that IRc4 is self-luminous because they determined that it coincides with the peak of the hot core defined by high-lying ammonia lines, and Aitken et al. (1981) suggested that IRc4 contains a luminosity source because of the shape of its deep silicate feature, which requires a range of dust temperatures. However, Wynn-Williams et al. (1984), and most investigators since (e.g., Gezari et al. 1998) have considered that IRc3 and IRc4 are externally heated knots of dust or are ``holes'' in the clumpy, dense shell surrounding a luminous cavity. Simpson et al. (2006) used the NICMOS polarimeter to show that IRc4 and IRc3 are highly polarized and they therefore argued that they are apparently illuminated by source \textit{I} and/or IRc2 (see also Werner et al. 1983). Okumura et al. (2011) produced a 7.8/12.4~$\mu$m color temperature map of the area and show there is a local minimum in the temperature at the location of IRc7, opposite what one would expect from an internally heated source.

Even though IRc7 becomes much more prominent at longer wavelengths than nearby IRc2, there are no clear temperature peaks in the FORCAST data at these two locations. The color temperature maps do show a region of relatively high temperature at their locations, and if they indeed contain secondary temperature peaks, they are not resolved in the maps.

On the other hand, the FORCAST data show that IRc4 is likely internally heated, because it peaks in intensity at 19.7, 31.5, and 37.1~$\mu$m at the same location as the peak in the 19.7/31.5~$\mu$m and 31.5/37.1~$\mu$m color temperature maps. If IRc4 were externally heated or were a ``hole'', there would be no temperature peak here and one would expect that the peak location would change with temperature (and hence wavelength observed), but we do not observe this behavior. We can also dismiss this external heating argument by calculating the amount of heating IRc4 encounters from source \textit{I} using the luminosity of source \textit{I} ($\sim$10$^5$~L$_{\sun}$ under the assumption it is responsible for all the emission of the KL nebula), along with the measured FWHM of IRc4 (4.2$\arcsec$), and the measured distance between the location of source \textit{I} and IRc4 (6$\arcsec$, which under the assumption they are both in the plane of the sky yields their \textit{minimum} separation). Assuming source \textit{I} radiates into 4$\pi$ steradians (a likely assumption if the claim is that all of the KL nebula is heated by this source), IRc4 encounters $\lesssim$3\% of the luminosity of source \textit{I}. This means that IRc4 would only intercept $\lesssim$3$\times$10$^3$~L$_{\sun}$, an order of magnitude lower than the value we derive.

In the case of IRc3, the peak flux density location moves as a function of wavelength in the images. Although this is consistent with its being an externally heated source, as most previous investigators have concluded, it peaks farther to the west at 19.7~$\mu$m, and further to the east at longer wavelengths. This is the opposite of what would be expected if the source were externally heated by source \textit{I}, the oft-presumed main luminosity source in the IRc2 region. One possible explanation for this would be if there were multiple internal heating components in the IRc3 region with different temperatures causing the peaks to occur at different positions at different wavelengths. Higher resolution maps in the mid-infrared (Shuping et al. 2004, Okumura et al. 2011) clearly show IRc3 resolved into at least 2 separate components (named IRc3N and IRc3S) at 12.5~$\mu$m, and that there is another tongue of emission and possibly another peak just west of IR3N. 
It is therefore plausible that there are multiple sources in this region responsible for the behavior seen in the data.

\medskip

\subsection{SOF1}

There is a source of emission in the area that has never been identified prior to this work. It can be seen best in the deconvolved 31.5 and 37.1~$\mu$m data (Figure 1). Marked on that figure as ``SOF1'', it appears not as a discrete source, but as an extended area of emission. It is present in both the natural resolution 31.5 and 37.1~$\mu$m data of Figure 1 as well, visible as the ``tongue'' of emission extending to the northeast from the IRc11/IRc2 region. There is also a hint of this emission at shorter wavelengths; for instance there is a bulge in this direction from IRc13 in the 20~$\mu$m images of Gezari et al. (1998), and a discrete yet faint source seen in the 12.5~$\mu$m Keck image of Shuping et al. (2004). There is also an enhancement of flux here in the 30 $\mu$m contours of Wynn-Williams et al. (1984). The emission is too distant from any other identified source to merely be a shifted peak within an already known source, indicating that it is a previously unidentified region of emission.

While the exact nature of this source is unclear, it is wholly contained within the northeastern lobe of the SiO outflow mapped by Plambeck et al. (2009), just as IRc3, IRc4, IRc5, and IRc7 are contained within the southwestern SiO outflow lobe. Therefore, it is plausible that these longer wavelengths permit one to peer through the overlying extinction, allowing detection of the thermal emission from the northeastern outflow cavity wall for the first time.

\acknowledgments

We would like to thank the FORCAST engineering team of George Gull, Justin Schoenwald, and Charles Henderson for their unwavering efforts to make FORCAST a reality. RYS is supported by USRA contract 209000771 to the Space Science Institute. Based on observations made with the NASA/DLR Stratospheric Observatory for Infrared Astronomy (SOFIA).  SOFIA science mission operations are conducted jointly by the Universities Space Research Association, Inc. (USRA), under NASA contract NAS2-97001, and the Deutsches SOFIA Institut (DSI) under DLR contract 50 OK 0901.

{\it Facilities:} \facility{SOFIA (FORCAST)}.


\begin{deluxetable}{lccccccccc}
\tabletypesize{\tiny}
\tablecaption{Orion BN/KL Integrated Source Flux Densities in Janskys}
\tablewidth{0pt}
\tablehead{
\colhead{Source} & \colhead{Position$^a$} & \colhead{Aperture$^b$} & \colhead{} &
\colhead{6.6$\mu$m} & \colhead{7.7$\mu$m} & \colhead{19.7$\mu$m} & \colhead{31.5$\mu$m} & \colhead{37.1$\mu$m} \\
}
\startdata
BN 		&0,0		&3.4$\times$3.4 &	&211		&134	 	 &490	 &1680	&2610   \\
IRc2	&+5.7,-6.8	&3.4$\times$3.4	&	&22.0	 	&33.3	 	 &243	 &2040	&3160   \\
IRc3	&-2.0,-7.6	&3.4$\times$3.4	&	&7.9		&7.8		 &451	 &2480	&3860   \\
IRc4	&+1.5,11.8	&4.3$\times$4.0	&	&11.3		&14.7		 &747	 &4280	&6520  \\
IRc6	&+0.1,-4.2	&3.0$\times$3.0	&	&22.7		&15.6		 &282	 &1690	&2700	  \\
IRc7	&+2.8,-8.0	&3.4$\times$2.5	&	&13.2	 	&17.1		 &287	 &1970	&3000	  \\
SOF1    &+11.8,-3.9 &4.0$\times$4.0 &	&3.0		&4.6		 &166	 &1680	&2800	  \\
BN/KL	&+5.7,-6.8	&32.0$\times$32.0 & &666		&571		 &12500 &81700	&143000   \\
\enddata
\tablecomments{Source flux densities are given in Jy at the indicated wavelengths in microns and were calculated summing the observed brightnesses in the given apertures in the calibrated (but not deconvolved) images with no corrections for any underlying extended environmental emission. These values are believed to have errors less than 20\%, except for the 6.6~$\mu$m data which is thought to have an error of 25\%.}
\tablenotetext{a}{Relative peak position ($\Delta$$\alpha$, $\Delta$$\delta$) compared to the peak of BN in arcseconds. For IRc3, IRc4, IRc6, and IRc7 the offsets given were measured from the 19.7$\mu$m data shown in Figure 1. For IRc2, the offsets come from Gezari et al. 1998.}
\tablenotetext{b}{Aperture sizes are square with dimensions given in arcseconds.}
\end{deluxetable}


\begin{figure}
\includegraphics[scale=.85]{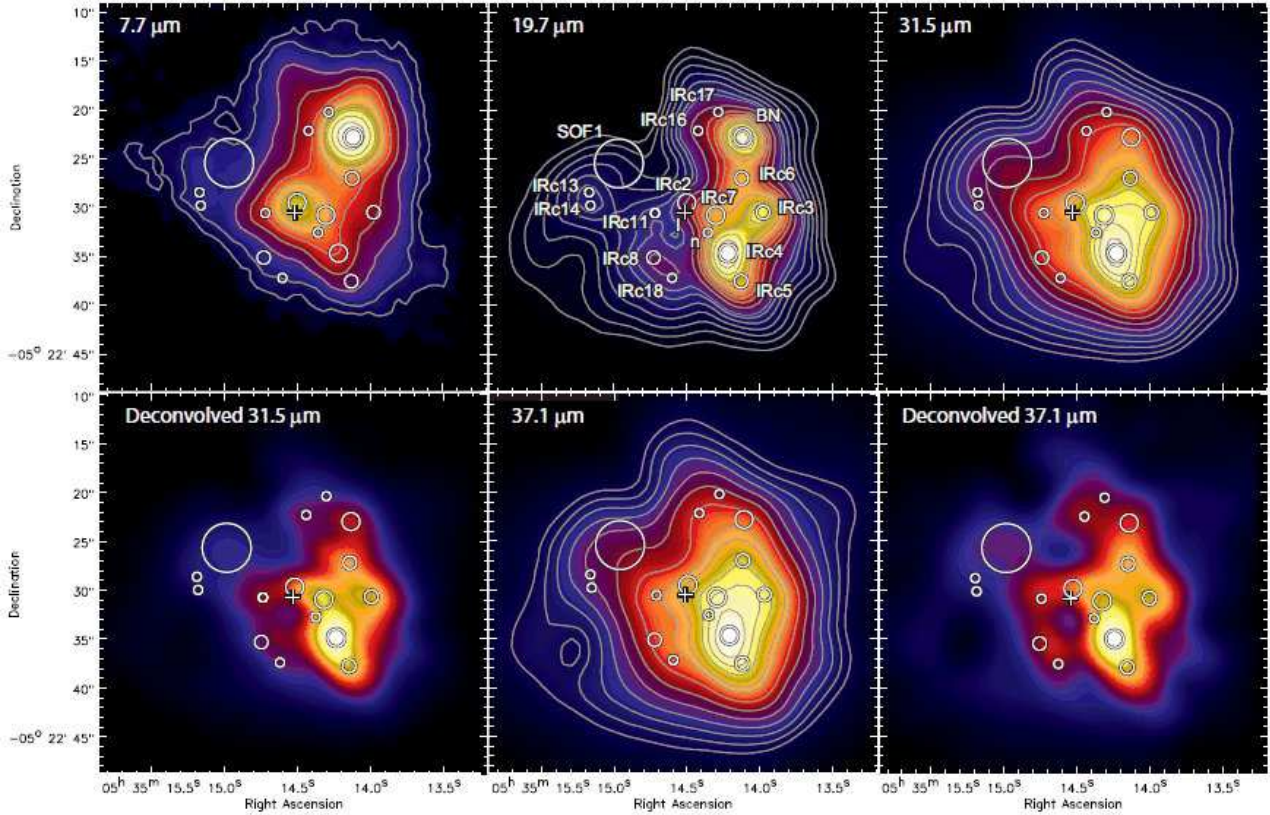}
\caption{The Orion BN/KL region at 4 wavelengths: 7.7, 19.7, 31.5, and 37.1~$\mu$m as taken with the FORCAST instrument on SOFIA. Also shown are the deconvolved 31.5, and 37.1 ~$\mu$m images. In the 19.7~$\mu$m panel, IR source positions are shown with open circles and labeled. These positions are given in Table 1 as offsets from the peak of BN at 19.7~$\mu$m. Radio source I is also labeled and its position is given by a cross. For cross-wavelength comparisons, in the remaining three panels the symbols only are also shown. Contours levels are: 0.10, 0.13, 0.21, 0.34, 0.55, 0.90, 1.45, 2.36, 4.86, 6.19 Jy/pixel at 7.7~$\mu$m; 3.72, 4.40, 5.20, 6.15, 7.27, 8.60, 10.2, 12.0, 14.2, 15.9, 17.8, 19.9, 22.2, 24.9, 27.8 Jy/pixel at 19.7~$\mu$m; 20.4, 24.0, 28.2, 33.2, 38.9, 45.8, 53.8, 63.3, 74.3, 87.4, 103, 120, 131, 142, 154 Jy/pixel at 31.5~$\mu$m; and 39.4, 47.3, 56.7, 68.0, 81.5, 97.5, 117, 141, 154, 185, 202, 221 Jy/pixel at 37.1~$\mu$m.}
\end{figure}

\begin{figure}
\includegraphics[scale=.5]{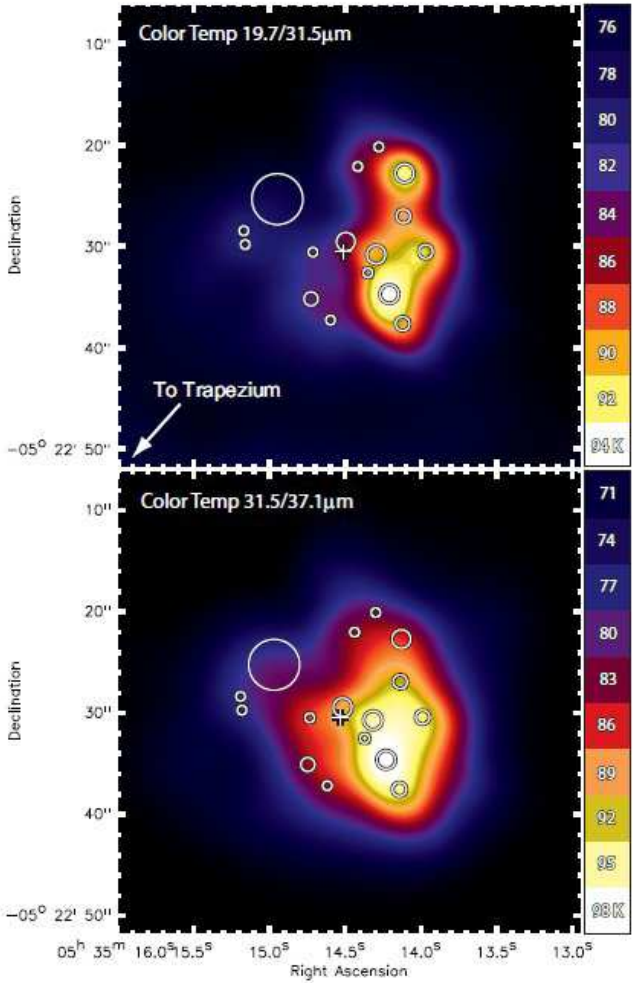}
\caption{Dust color temperature maps of Orion BN/KL. The top panel shows the map made from the 19.7 and 31.5~$\mu$m images, and the bottom shows the map made from the 31.5 and 37.1~$\mu$m images. All symbols are as described in Figure 1. }
\end{figure}

\begin{figure}
\includegraphics[scale=0.4]{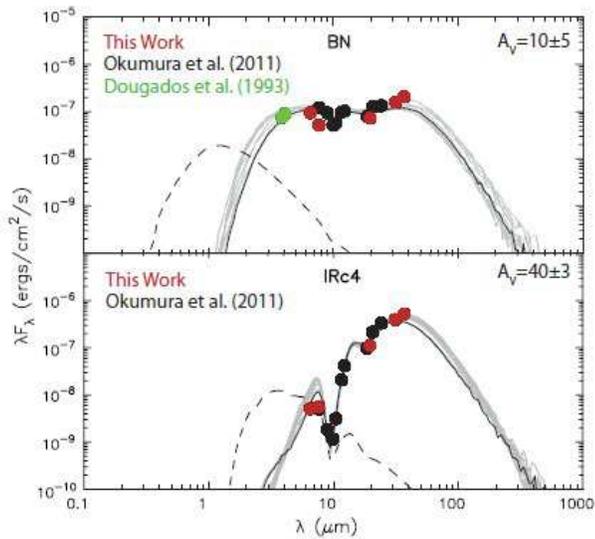}
\caption{SED models (from Robitaille et al. 2007) fit to the SOFIA data and data in the literature for BN (top) and IRc4 (bottom). The top ten best model fits are shown in gray, with the black line showing the best fit. The dashed line shows the stellar photosphere corresponding to the central source of the best fitting model, as it would look in the absence of circumstellar dust (but including interstellar extinction).}
\end{figure}

\end{document}